\documentclass[superscriptaddress,amsmath,amssymb, aps, pra,twocolumn]{revtex4-2}

\usepackage{graphicx}
\usepackage{dcolumn}
\usepackage{bm}
\usepackage[colorlinks]{hyperref}
\usepackage{amsthm}
\usepackage[dvipsnames]{xcolor}
 
\newcommand{\be}{\begin{equation}}
\newcommand{\ee}{\end{equation}}
\newcommand{\beq}{\begin{eqnarray}}
\newcommand{\eeq}{\end{eqnarray}}

\newcommand{\ket}[1]{\mbox{$ | #1 \rangle $}}
\newcommand{\bra}[1]{\mbox{$ \langle #1 | $}}

\newcommand{\expval}[1]{\mbox{$\langle #1 \rangle$}}

\newcommand{\orcid}[1]{\href{https://orcid.org/#1}{\includegraphics[width=7pt]{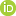}}}

\begin{document}

\title{Thermal devices powered by generalized measurements with indefinite causal order}

\author{Pedro R. Dieguez\orcid{0000-0002-8286-2645}}\email{dieguez.pr@gmail.com}
\affiliation{International Centre for Theory of Quantum Technologies, University of Gdańsk, Jana Bazynskiego 8, 80-309 Gdańsk, Poland.}
\affiliation{Center for Natural and Human Sciences, Federal University of ABC, Avenida dos Estados 5001, 09210-580, Santo Andr\'{e}, S\~{a}o Paulo, Brazil.}

\author{Vinicius F. Lisboa}
\affiliation{Center for Natural and Human Sciences, Federal University of ABC, Avenida dos Estados 5001, 09210-580, Santo Andr\'{e}, S\~{a}o Paulo, Brazil.}

\author{Roberto M. Serra\orcid{0000-0001-9490-3697}}\email{serra@ufabc.edu.br}
\affiliation{Center for Natural and Human Sciences, Federal University of ABC, Avenida dos Estados 5001, 09210-580, Santo Andr\'{e}, S\~{a}o Paulo, Brazil.}
\affiliation{Department of Physics, Zhejiang Normal University, Jinhua 321004, China}

\begin{abstract}
A quantum-controlled device may produce a scenario in which two general quantum operations can be performed in such a way that it is not possible to associate a definite order for the operations application. Such an indefinite causal order can be explored to produce nontrivial effects in quantum thermal devices. We investigate a measurement-powered thermal device that consists of generalized measurement channels with adjustable intensity parameters, where energy is exchanged with the apparatus in the form of work or heat. The measurement-based device can operate as a heat engine, a thermal accelerator, or a refrigerator, according to a measurement intensity setting. By employing a quantum switch of two measurement channels, we explore a thermal device fueled by an indefinite causal order. We also discuss how a coherent control over an indefinite causal order structure can change the operating regimes of the measurement-powered thermal device to produce an advantage when compared to a scenario with an incoherent control of the order switch.
\end{abstract}

\maketitle

\section{Introduction}

A systematic understanding of the superposition principle and its role applied to quantum systems and quantum-controlled processes as well are essential to a deeper understanding of many foundational and applied questions such as the indefiniteness of realism in quantum systems~\cite{dieguez2018information,lustosa2020irrealism}, wave-and-particle duality~\cite{PhysRevA.85.032121,adesso2012,dieguez2022experimental}, thermodynamic arrow of time~\cite{micadei2019,PhysRevE.97.062105,rubino2021quantum}, and quantum gravity~\cite{hardy2007towards,zych2019bell}. The quantum-controlled switch of the application order of two or more quantum maps showed new counter-intuitive predictions concerning the possibility to have an indefinite causal order~\cite{chiribella2013quantum,oreshkov2012quantum,brukner2014quantum,araujo2015witnessing,rubino2017experimental,oreshkov2019time,goswami2018indefinite,goswami2020experiments}. Processes with indefinite causal order were explored with several novel advantages and applications for quantum computation \cite{chiribella2013quantum, araujo2014computational, procopio2015experimental}, communication~\cite{ebler2018enhanced, wei2019experimental, guo2020experimental,goswami2020increasing,rubino2021experimental,guerin2019communication}, metrology~\cite{zhao2020quantum},  and quantum thermodynamics~\cite{felce2020quantum, rubino2021quantum, cao2022quantum,nie2022experimental,felce2021refrigeration,simonov2022work}.

Quantum thermodynamics~\cite{kosloff2013quantum,vinjanampathy2016quantum, Goold_2016} emphasized the relevance of thermal and quantum fluctuations~\cite{esposito2009nonequilibrium,campisi2011colloquium,batalhao2014experimental,PhysRevLett.115.190601,arxiv.2104.13427} for proper manipulation of the system's energetic \cite{arxiv.2111.09241}. The recent development of quantum thermal devices can find possible applications in the emerging downscale technologies~\cite{myers2022quantum,campisi2011colloquium,batalhao2014experimental,campisi2014fluctuation,elouard2015reversible,campisi2015nonequilibrium,dechant2015all, peterson2019experimental,klatzow2019experimental, altintas2015general,  rossnagel2016single, barontini2019ultra, bouton2021quantum}.
A new class of thermal devices powered by quantum measurements was introduced, and they were realized by exploring the fact that measurements on a quantum system are invasive and may change its internal energy~\cite{yi2017single,elouard2017role,brandner2015coherence}. Employing non-selective measurements, a single temperature heat engine without feedback control to extract work~\cite{yi2017single} was recently introduced. In addition, several measurement-based thermodynamic protocols were studied in the last few years~\cite{campisi2017feedback,chand2017single,chand2017measurement,mohammady2017quantum,elouard2017extracting,chand2018critical,ding2018measurement,elouard2018efficient,buffoni2019quantum,solfanelli2019maximal,jordan2020quantum,behzadi2020quantum,seah2020maxwell,bresque2021two,chand2021finite,lin2021suppressing,PhysRevLett.124.110604, PhysRevLett.121.030604, PRXQuantum.3.020329,
PhysRevLett.117.240502, anka2021measurement,alam2022two,manikandan2022efficiently,myers2022quantum,lisboa2022experimental}. Recently, a proof-of-concept experiment of a spin heat engine powered by generalized measurements employing a nuclear magnetic resonance platform~\cite{lisboa2022experimental} was performed. The cycle consists in two non-selective generalized measurements channels with adjustable measurement strengths, one dedicated to fueling the device, then playing the role of a heat source, and the other committed to work extraction when applied in an isentropic way. By changing the internal energy in an isentropic way, the working substance is considered informationally closed in a way that this operation can be recognized as work extraction. An interesting point, which was observed in the experimental results reported in Ref.~\cite{lisboa2022experimental}, is that this kind of quantum thermal device can reach unit efficiency while also achieving maximum extracted power at the same time with the fine-tuning of the measurement strengths. 

Thermodynamic tasks, for instance, cooling quantum systems, are pivotal to many applications in quantum technologies. Considering further applications, it is of significant importance to look for methods to perform such thermodynamic tasks on quantum devices in fast and efficient ways. Interestingly, the intrinsic indefiniteness in the causal structure revealed by a quantum-controlled process (with indefinite causal order) was recently employed to perform a refrigerator cycle~\cite{felce2020quantum,felce2021indefinite}, which suggests a new kind
of non-classical resource for a thermodynamic task~\cite{felce2020quantum,rubino2021quantum}.
A quantum-controlled switch of two processes was present in an investigation of a possible indefinite thermodynamic arrow of time~\cite{rubino2021quantum}. It was demonstrated how the entropy of a quantum system can distinguish between two directions (forward and backward evolution) of a mutually time-reversal thermodynamic processes~\cite{rubino2021quantum}. This was done by extending the so-called two-point measurements (TPM)~\cite{kosloff2013quantum,vinjanampathy2016quantum,esposito2009nonequilibrium} protocol to encompass indefinite causal order structures to analyze their consequences on the stochastic work distribution performed on each thermodynamic evolution (in the forward and backward direction). Moreover, it was demonstrated how quantum interference effects can be used to
reduce undesired thermal fluctuations resulting from an effective quantum-controlled superposition of a heat engine and a power-driven refrigeration~\cite{rubino2021quantum}.

In this work, we first present a non-selective generalized measurement powered cycle that can operate in the thermal accelerator, heat engine, or refrigerator modes, by just adjusting the generalized measurement parameters. The cycle is composed of two measurement channels, one playing the role of heat source and another dedicated to invest or extract work. By employing information-theoretical quantities, we show the impossibility to extract work directly from a thermal equilibrium state using an isentropic non-selective generalized measurement channel. Thus, it is necessary to pass through one or more non-equilibrium states before extracting work using a non-selective generalized measurement channel. In the following, we employ a quantum-controlled switch of the two causal orders in which the measurement channels are being applied to explore unusual quantum interference effects on the performance of thermal devices. More specifically, the device operation regime is extended to encompass a parameter's set in which it does not operate when the causal order of the measurement channels is defined.  The difference between a coherent and incoherent control of the order switcher will also be explored.


\section{Thermal devices powered by generalized measurements}

Quantum heat engines allow work extraction from a heat flow from hot to cold thermal baths, while thermal accelerators reinforce this natural heat flow by investing energy in the form of work. On the other hand, quantum refrigerators have the purpose of pumping heat from a cold to a hot environment by investing work. 

The quantum thermal devices proposed here perform cycles with the use of only one temperature reservoir which will be considered as a cold environment, and a set of non-selective generalized measurements that will act upon a two-level working or cooling substance. 
The generalized measurements are described as completely positive and trace-preserving (CPTP) maps with varying measurement parameters. It can be implemented in experimental contexts by building suitable positive operator-valued measurements (POVMs), as performed, for instance, in Ref.~\cite{lisboa2022experimental} in a
proof-of-concept experiment using nuclear magnetic resonance techniques to investigate a spin
quantum heat engine driven by non-selective weak measurements without feedback
control. The specific outcome of the generalized measurement is associated with a given probability, and to have a deterministic result of the cycle, we also consider here two non-selective (without post-selection) generalized measurements channels with suitable adjusted parameters, with the only difference that instead of working with the generalized measurements investigated in Ref.~\cite{lisboa2022experimental}, we change their structure to make easy the further analysis for the reverse order and quantum switch protocols. A measurement apparatus can be modeled as an device that has some microscopic quantum degrees of freedom which is further amplified to the classical realm through a macroscopic pointer that is coupled with the former~\cite{lisboa2022experimental}. Along the generalized measurement thermal cycle, energy is exchanged with two different meters via the interaction of the system with the meter's internal degrees of freedom. When the measurement channel changes the von Neumann entropy of the working or cooling substance, the stochastic energy exchanged with the meter is recognized as heat. On the other hand, when the von Neumann entropy remains constant along the measurement channel (isentropic channel), the energy exchanged with the meter has a nature of work~\cite{behzadi2020quantum,alipour2022entropy,lisboa2022experimental}.

%
\begin{figure}
    \centering
    \includegraphics[width=1\columnwidth]{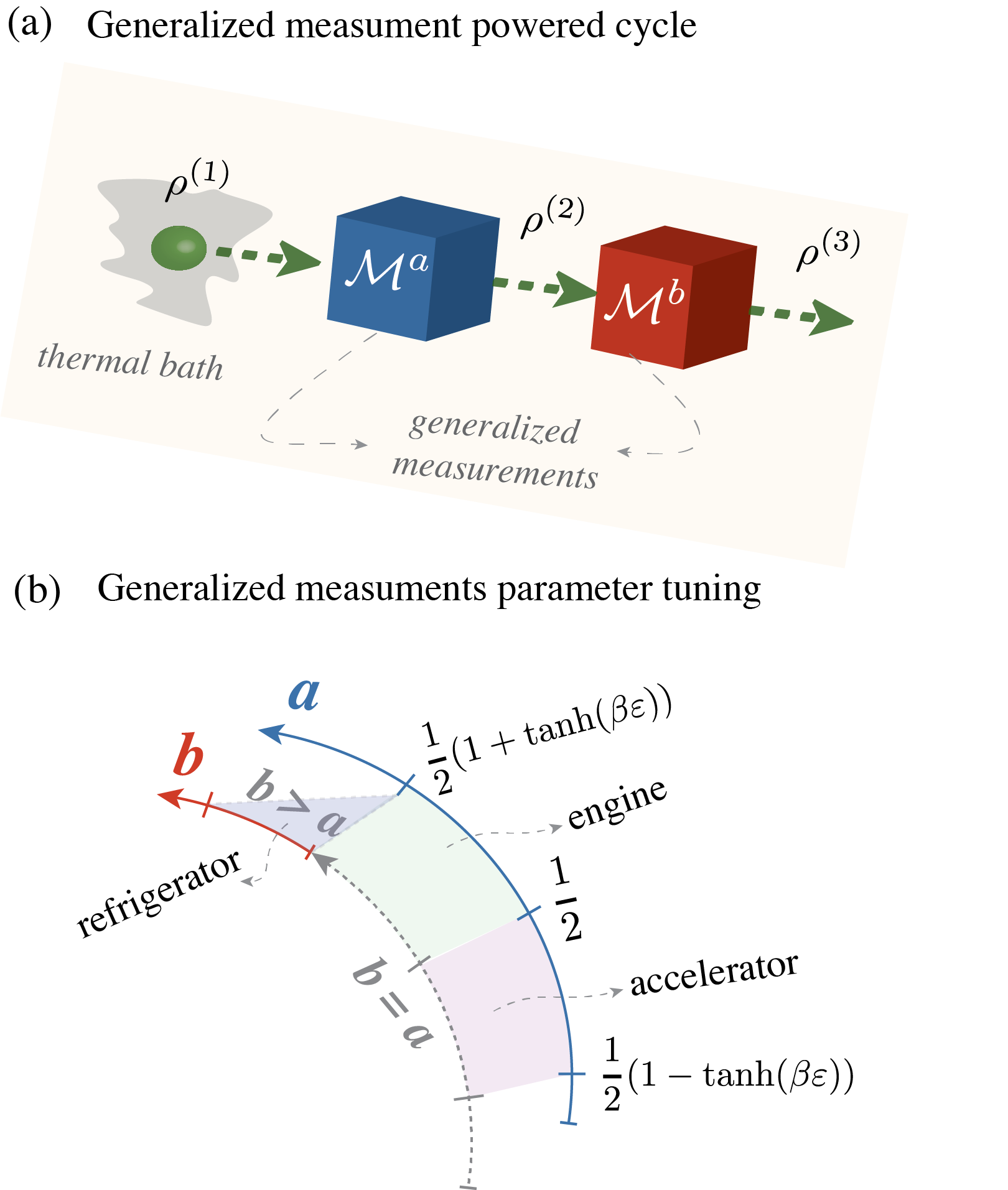}%
    \caption{ Generalized measurements powered thermal device. (a) Schematic representation of the quantum thermal cycle powered by generalized measurements. In the first stroke, the system is initialized in the cold thermal equilibrium state $\rho^{(1)}$ at inverse temperature $\beta$. The second stroke is a generalized measurement channel $\mathcal{M}^a$ with a varying measurement (strength) parameter $a$. The third stroke is another generalized measurement $\mathcal{M}^b$ with a varying measurement parameter $b$. Both measurement channels represent non-selective measurements. (b) Description of the measurement parameter setting to perform the measurement powered cycle in three operation modes: accelerator, engine, or refrigerator.}
    \label{fig:1}
\end{figure}
\begin{figure}
    \centering
    \includegraphics[width=1\columnwidth]{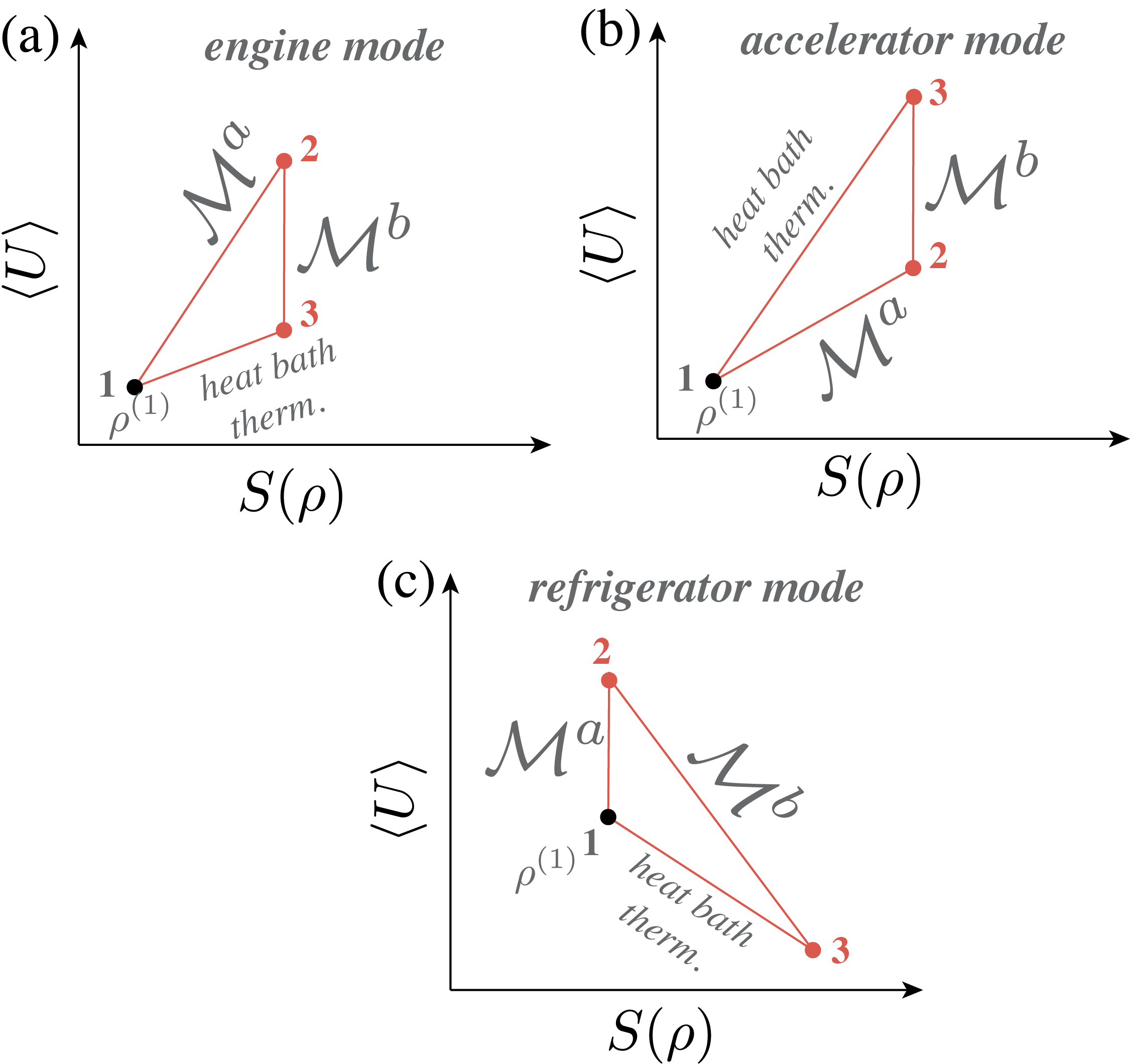}%
    \caption{ Thermal devices powered by non-selective generalized measurements (a) Engine cycle. Straining from a thermalization with the cold environment two generalized measurement channels are sequentially applied. The $\mathcal{M}^a$ channel increases the energy and von Neumann entropy of the working substance (heat absorption from the meter $\mathcal{A}$), while the second measurement $\mathcal{M}^b$ is dedicated to work extraction (to the meter $\mathcal{B}$) by changing the internal energy in an isentropic way. (b) Thermal accelerator cycle, with the first measurement parameter adjusted (in the meter $\mathcal{A}$) to increase energy and entropy of the working substance, while the second measurement channel adds energy in an isentropic way (work invested by the meter $\mathcal{B}$). (c) Refrigerator cycle, with a first isentropic measurement channel (meter $\mathcal{A}$) which performs work on the system, while the second measurement parameter is now adjusted to produce a heat flux from the cold environment to the meter $\mathcal{B}$. }
    \label{fig:2}
\end{figure}

\subsection{Thermal devices operated in the forward order}\label{IIB}

The generalized measurements powered cycle consists of three strokes, depicted in Fig.~\ref{fig:1}(a). We consider a single-qubit working or cooling substance, however the ideas presented here can be generalized to higher dimensional systems. The first stroke is a full thermalization with a cold environment, so that the working or cooling substance will be initialized in a thermal (Gibbs) equilibrium state, $\rho^{(1)}=\text{exp}\left[-\beta H\right]/\mathcal{Z}$,
at  inverse spin temperature $\beta=1/\left(k_{B}T\right)$, with
$\mathcal{Z}=\text{tr}\left[\text{exp}\left(-\beta H\right)\right]$ as the associated partition function, and a generic Hamiltonian $H=-\varepsilon\sigma_z$ (where $\sigma_{(x,y,z)}$ are the usual Pauli matrices and $2\varepsilon$ is the energy gap) that will be fixed along the cycle.

In the second stroke, the working or cooling substance undergoes a non-selective measurement channel, whose effect can be described by a CPTP map with a Kraus decomposition, $\mathcal{M}^a:\rho^{(1)}\rightarrow \rho^{(2)}=\sum_jM^a_j \rho^{(1)} M^{a\dagger}_j$. For this channel, we choose the following Kraus operators $M^a_1=\sqrt{1-a}\ket{0}\bra{0}$, $M^a_2=\sqrt{1-a}\ket{0}\bra{1}$, $M^a_3=\sqrt{a}\ket{1}\bra{1}$, and $M^a_4=\sqrt{a}\ket{1}\bra{0}$, with measurement parameter $a$. The POVM associated to $j$-th generalized measurement outcome is $M^{a\dagger}_j M^a_j$ (occurs with probability $\text{tr}(M^{a\dagger}_j M^a_j\rho^{(1)})$) and satisfies $\sum_{j}M^{a\dagger}_j M^a_j=\openone$. After the application of the measurement channel, the working substance evolves to the out of equilibrium state (with relation to the cold environment) $\rho^{(2)}=\mathcal{M}^a(\rho^{(1)})$. In this stage, the meter $\mathcal{A}$ leads to a change in the system internal energy given by
 \begin{equation} \label{U2}
 \begin{aligned}
      \expval{\Delta U^{(2)}}&=\text{tr}\left[H\left(\rho^{(2)}-\rho^{(1)}\right)\right] \\
      &=2\varepsilon\left[a-\frac{1}{2}\left(1-\tanh(\beta \varepsilon)\right)\right] .
\end{aligned}
\end{equation}

The von Neumann entropy variation $\Delta S^{(2)}=S(\rho^{(2)})-S(\rho^{(1)})$ (with $S(\rho)=-\text{tr}\rho \ln \rho$) for the  measurement channel $\mathcal{A}$ can be written as function of the binary entropy, $h(u):=-u \ln(u)- (1-u) \ln(u)$, as 
\begin{equation}
    \Delta S^{(2)}= h(a)-h\left( \frac{1}{2}(1-\tanh(\beta \varepsilon))\right).
\end{equation}

It is straightforward to note that for $a=\frac{1}{2}\left(1\pm\tanh(\beta \varepsilon)\right)$ this measurement channel  is isentropic, and consequently, the energy exchange with meter occurs in form of work. Meanwhile, for $\frac{1}{2}\left(1-\tanh(\beta \varepsilon)\right)< a < \frac{1}{2}\left(1+\tanh(\beta \varepsilon)\right)$ it has a positive entropy variation, $ \Delta S^{(2)}>0$. In this case the meter will supply energy in form of heat to the system.  

In the third stroke, another measurement channel will be applied  $\mathcal{M}^b:\rho^{(2)}\rightarrow \rho^{(3)}=\sum_jM^b_j \rho^{(2)} M^{b\dagger}_j$ with the Kraus operators 
 $M^b_1=\sqrt{1-b}\ket{1}\bra{1}$, $M^b_2=\sqrt{1-b}\ket{1}\bra{0}$, $M^b_3=\sqrt{b}\ket{0}\bra{0}$ and $M^b_4=\sqrt{b}\ket{0}\bra{1}$. In this stroke the associated POVM is $M^{b \dagger}_j M^b_j$ with $\sum_{j}M^{b \dagger}_j M^b_j=\openone$. After the measurement channel $\mathcal{B}$, the system will be in the state
$\rho^{(3)}=\mathcal{M}^b\left(\mathcal{M}^a(\rho^{(1)})
\right)$.   
In this stage, the mean internal energy variation of the system due to the measurement channel and the entropy variation are  
\begin{equation}  
\label{U3}
    \expval{\Delta U^{(3)}}=\text{tr} \left[H\left(\rho^{(3)}-\rho^{(2)}\right)\right]=2\varepsilon(1-a-b) ,
\end{equation}
and 
\begin{equation}
    \Delta S^{(3)}= h(b)-h(a),
\end{equation}
respectively.

To close the cycle, the working or cooling substance will interact again with the cold environment at inverse temperature $\beta$. The system's internal energy change (heat exchange with the cold environment) is given by
 \begin{equation}\label{U1}
 \begin{aligned}
    \expval{ \Delta U^{(1)}}&=\text{tr}\left[H\left(\rho^{(1)}-\rho^{(3)}\right)\right] \\
    &= -2\varepsilon\left[\frac{1}{2}\left(1+\tanh(\beta \varepsilon)\right)-b\right] .
 \end{aligned}
 \end{equation}
 
Depending on the choices of the measurement channel parameters $a$ and $b$, the cycle can act as an engine, a thermal accelerator, or a refrigerator. 
First, if we adjust the measurement channel $\mathcal{B}$ interaction parameter to have the same intensity as in the channel $\mathcal{A}$, i.e. $b=a$, it leads to an isentropic process in the stroke three, $\Delta S^{(3)}=0$, but with a change in the internal energy given by, $\expval{\Delta U^{(3)}}=2\varepsilon(1-2a)$ and related to a work absorbed or delivered from or to the measuring apparatus $\mathcal{B}$. 

To operate the thermal quantum device in the accelerator and heat engine modes, we should have in the stroke two, $\expval{\Delta U^{(2)}}=\mathcal{Q}_{\text{hot}}>0$ equivalent to a flow of stochastic energy from the meter $\mathcal{A}$ to the system. For the thermal accelerator, the measurement parameter of the channel $\mathcal{A}$ is limited by $\frac{1}{2}\left(1-\tanh(\beta \varepsilon)\right)< a < \frac{1}{2}$ and for the engine mode it is constrained as 
$\frac{1}{2}\leqslant a < \frac{1}{2}\left(1+\tanh(\beta \varepsilon)\right)$. Such intervals combined with parameter choice $b=a$  in the measurement channel $\mathcal{B}$, result in $\expval{\Delta U^{(3)}}=\mathcal{W}>0$ in the accelerator mode, and $\expval{\Delta U^{(3)}}=\mathcal{W}\leqslant0$ for the engine mode. In both operation modes, $\expval{\Delta U^{(1)}}=\mathcal{Q}_{\text{cold}}<0$, will be the heat delivered to the thermal bath to close the cycle.  

The refrigerator mode is obtained when we explore the first measurement channel ($\mathcal{A}$) as a source of thermodynamic work, adjusting the measurement parameter as $a=\left(1+\tanh(\beta \varepsilon)\right)/2$.
In this case $\mathcal{W}=\expval{\Delta U^{(2)}}=2\varepsilon\tanh(\beta \varepsilon)>0$ and the channel $\mathcal{A}$ turns out to be an isentropic one, $\Delta S^{(2)}=0$. In the next stages a heat flow from the cold environment to the meter $\mathcal{B}$ is implemented by adjusting $b>\frac{1}{2}\left(1+\tanh(\beta \varepsilon)\right)$, which leads to $\expval{\Delta U^{(3)}}=\mathcal{Q}_{\text{hot}}<0$,  $\expval{\Delta U^{(1)}}=\mathcal{Q}_{\text{cold}}>0$, and $\Delta S^{(3)}>0$. In this regime a heat flux form the cold environment to the meter $\mathcal{B}$ will be established.       

The possible measurement parameters choices and the operation modes of the quantum thermal device are illustrated in Fig.~\ref{fig:1}(b). The operation mode of each generalized measurement powered cycle is also sketched in Fig.~\ref{fig:2}(a) to \ref{fig:2}(c) in terms of the internal energy and von Neumann entropy variations.

Before commenting on the figures of merit for each operation mode, let us rewrite Eqs.~(\ref{U2}), (\ref{U3}), and (\ref{U1}) in terms of entropic quantities. The internal energy variation in the $\ell$-th stroke can be written as  
\begin{equation}\label{relative}
\beta \expval{\Delta U^{(\ell)}}=\Delta S^{(\ell)}+ \Delta S(\rho^{(\ell)}||\rho^{(1)}),
\end{equation}
where we introduced the difference of relative entropy with respect to the thermal state, defined as $\Delta S(\rho^{(\ell)}||\rho^{(1)}) \equiv S(\rho^{(\ell)}||\rho^{(1)}) - S(\rho^{(\ell-1)}||\rho^{(1)})$ (for $\ell = 1$ we adopt $\rho^{(\ell-1)} =\rho^{(3)}$), $S(\rho^{(\ell)}||\rho^{(1)})=\text{tr}
\left[\left(\rho^{(\ell)} - \rho^{(1)}\right)\ln\rho^{(1)}\right]$ is the relative entropy, and the von Neumann entropy variation is $\Delta S^{(\ell)}\equiv S^{(\ell)} - S^{(\ell-1)}$ (for $\ell = 1$ we adopt $S^{(\ell-1)} =S^{(3)}$).

From Eq.~(\ref{relative}), we observe a general result that it is not possible to extract work directly from a thermal equilibrium state using only an isentropic non-selective measurement channel. To make this point clear, suppose that in the second stroke we perform a general isentropic map, so $\Delta S^{(2)}=0$ and the internal energy variation turns out to be  $\expval{\Delta U^{(2)}}=k_BT S(\rho^{(2)}||\rho^{(1)})\geq 0$, since the relative entropy is non-negative, work can only be inserted in the system. Therefore, to extract work from a non-selective generalized measurement protocol it is necessary to perform measurement channels that lead the system to an out-of-equilibrium state before applying the extracting work channel. Of course, using selective measurements or its combination with a feedback control protocol, it is possible to extract work from the system in a thermal state probabilistically \cite{koski2014experimental,aydiner2021quantum,davies2021harmonic}.  
It is interesting to emphasize at this point, that the present model of a thermal device powered by measurements is deterministic since we are considering the complete set of outcomes of the measurement without post-selection.

The figure of merit for the accelerator mode is the coefficient of performance ($\text{COP}^{\text{acc}}$) giving by the ratio between heat delivered to the cold environment ($\expval{\Delta U^{(1)}}=\mathcal{Q}_{\text{cold}}$) and the amount of work invested ($\expval{\Delta U^{(3)}}=\mathcal{W}$), that can be written as   
\begin{equation} \label{COPacc}
        \begin{aligned}
           \text{COP}^{\text{acc}}&=-
        \frac{\mathcal{Q}_{\text{cold}}} 
        {\mathcal{W}}=\frac{S(\rho^{(3)}||\rho^{(1)})+\Delta S_2}{S(\rho^{(3)}||\rho^{(1)})-S(\rho^{(2)}||\rho^{(1)})} \\
       &= \frac{1}{2} \left(1-\frac{\tanh(\beta \varepsilon)}{2a-1}\right).
        \end{aligned}
        \end{equation}

The efficiency in the heat engine mode is given by the amount of extracted work from the system by the meter $\mathcal{B}$ ($\mathcal{W}_{\text{ext}} = -\expval{\Delta U^{(3)}}$) divided by the amount of heat absorbed from the meter $\mathcal{A}$, 
\begin{equation} \label{eta}
   \begin{aligned}
       \eta&=\frac{\mathcal{W}_\text{ext}}{\mathcal{Q}_{\text{hot}}}=\frac{S(\rho^{(2)}||\rho^{(1)})-S(\rho^{(3)}||\rho^{(1)})}{S(\rho^{(2)}||\rho^{(1)})+\Delta S_2} \\
      &=  2 \left(1+\frac{\tanh(\beta \varepsilon)}{2a-1} \right)^{-1}.
   \end{aligned}
\end{equation}

In the refrigerator mode, the coefficient of performance ($\text{COP}^{\text{ref}}$) is defined as the ratio between the amount of heat absorbed from the cold environment and the work invested by the measurement channel  $\mathcal{M}^a$,
\begin{equation} \label{COPref}
       \begin{aligned}
            \text{COP}^{\text{ref}}&=  \frac{\mathcal{Q}_{\text{cold}}} 
        {\mathcal{W}}=\frac{\Delta S_1-S(\rho^{(3)}||\rho^{(1)})}{S(\rho^{(2)}||\rho^{(1)})} \\
           &=\left( b- \frac{1}{2} \right)\text{coth}(\beta\varepsilon)-\frac{1}{2}.
       \end{aligned}
\end{equation}

In addition to the microscopic informational analysis, the figures of merit for each operation mode written in terms of the entropy quantities, are useful for designing different measurement channels for each thermodynamic task.   

\subsection{Thermal devices operated in the backward order}

Before introducing the switch model for the two generalized non-selective measurements, first let us discuss the result of changing the causal order in which the maps ($\mathcal{A}$ and $\mathcal{B}$) are being applied. In other words, we consider in the following the reverse order of the cycle in Sec.~\ref{IIB}. It is straightforward from the symmetry of the maps, that changing the channel $\mathcal{A}$ by $\mathcal{B}$ is the same as using the following relation $b'=1-a$.
Then, it is easy to recover the engine, thermal accelerator, and refrigerator operating regimes with the reverse order of the sequential measurement channels,  $\mathcal{B}$ followed by $\mathcal{A}$. When carrying out the first measurement channel $\mathcal{M}^{b'}$ after the first thermalization, the average change in internal energy of the working substance is $\expval{\Delta U^{(2)}}=2\varepsilon\left[\frac{1}{2}\left(1+\tanh(\beta \varepsilon)\right)-b'\right]$. Note that, this value corresponds to Eq.~(\ref{U2}) with $b'=1-a$. In the third stroke, we consider now $\mathcal{M}^{a'}$ to have $\expval{\Delta U^{(3)}}=2\varepsilon(a'+b'-1)$, which can be identified with Eq.~(\ref{U3}) with $a'=1-b$ and  $b'=1-a$.  
The heat engine and thermal accelerator  can be obtained setting  $a'=b'$ in such way that the regimes are inverted in the reverse order, $\frac{1}{2}\left(1-\tanh(\beta \varepsilon)\right)< a' \leqslant \frac{1}{2}$ for thermal accelerator, and  $\frac{1}{2}< a' < \frac{1}{2}\left(1+\tanh(\beta \varepsilon)\right)$ for heat engine. Then, we can write $\text{COP}^{\text{acc}}= \frac{1}{2} \left(1+\frac{\tanh(\beta \varepsilon)}{2b'-1}\right)$ and  $\eta=  2 \left(1-\frac{\tanh(\beta \varepsilon)}{2b'-1} \right)^{-1}$. The refrigerator can be found by adjusting $a'< b'=\frac{1}{2}(1-\tanh(\beta \varepsilon))$ with $\text{COP}^{\text{ref}}=\left( \frac{1}{2}-a' \right)\text{coth}(\beta\varepsilon)-\frac{1}{2}$. In the following, we explore the employment of quantum switch to investigate the role of indefinite causal order in the measurement-based thermal devices.
 
\section{Quantum switch of the generalized measurement channels}

In this section, we examine the quantum switch of causal orders between the two measurement channels in the cycle described in the previous section. With that, we will be able to explore quantum interference associated with different application orders to enhance a thermal cycle based on generalized measurements. 

\begin{figure}
    \centering
    \includegraphics[width=1\columnwidth]{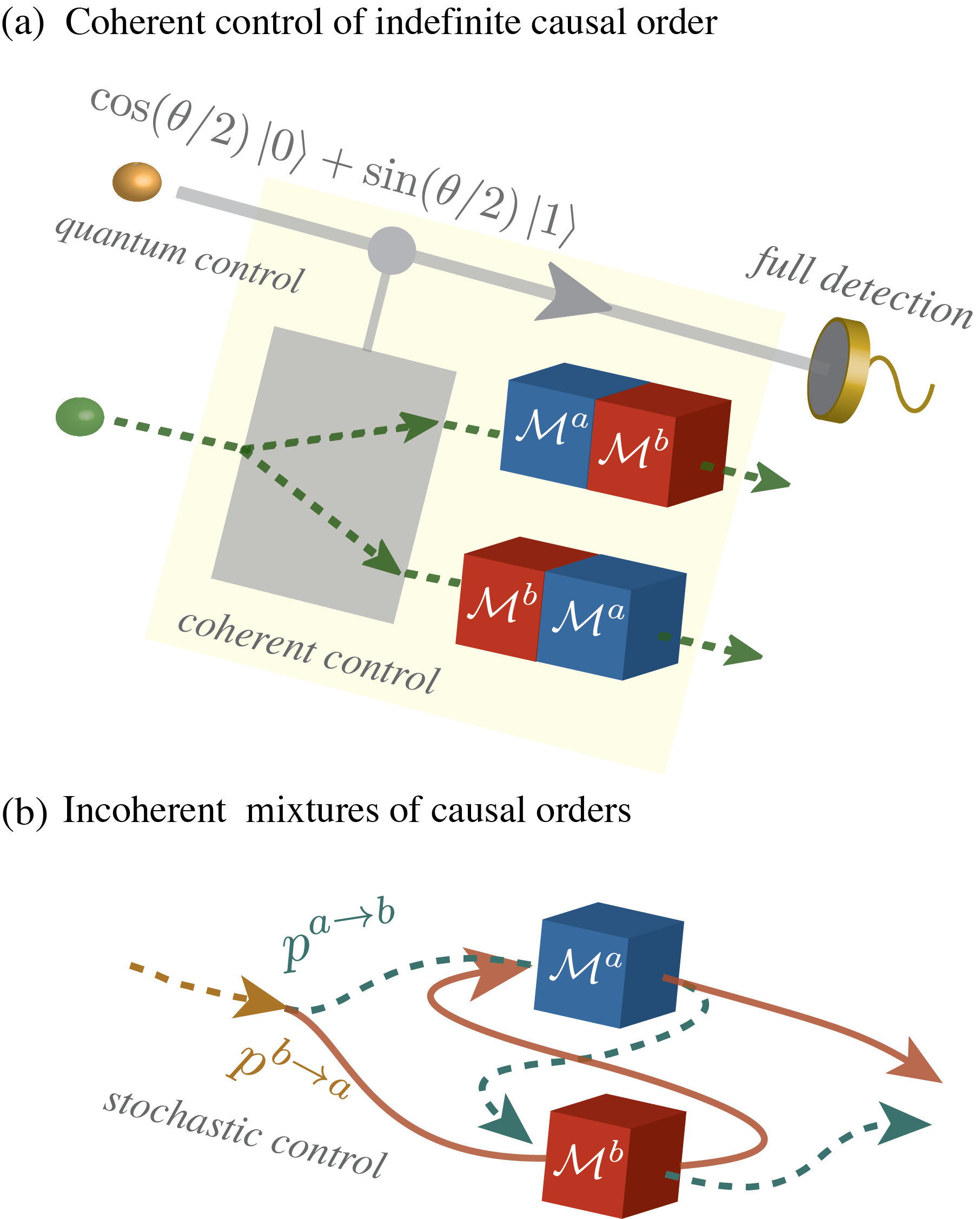}%
    \caption{ Indefinite causal order between measurement channels $\mathcal{M}^a$ and $\mathcal{M}^b$. (a) Coherent control of generalized measurement channels causal order which is realized by projecting the controller in an orthogonal basis which produces in each post-selected results an indefinite causal order. (b) Schematic description of the incoherent mixtures of causal orders which are implemented by tracing (non-observing) the control qubit. This is equivalent to a stochastic control of the two causal orders of the measurement maps, i.e., $\mathcal{M}^b\left(\mathcal{M}^a (\rho^{(1)})\right)$ with probability $p^{a\rightarrow b}$ and $\mathcal{M}^a\left(\mathcal{M}^b(\rho^{(1)})\right)$ with probability $p^{b\rightarrow a}$.}
    \label{fig:3}
\end{figure}

The non-classical switch of causal orders is realized by employing a quantum-controlled device, which here will be a control qubit parametrized as $\ket{c_{\theta}}=\cos{\frac{\theta}{2}}\ket{0}+\sin{\frac{\theta}{2}}\ket{1}$ ($0\leq \theta \leq \pi$). We choose to associate the control qubit state $\ket{0}$ with the sequential application of the channels $\mathcal{A}$ and $\mathcal{B}$ (as in the previous section), while the state $\ket{1}$ will be associated with the switched order of the channels application, as depicted in Fig.~\ref{fig:3}(a). When the controller has initial coherence, it is possible to obtain an indefinite causal order for the measurement maps application~\cite{felce2020quantum,simonov2022work}. Let us consider the case when the measurement parameters, of the channels $\mathcal{A}$ and $\mathcal{B}$,  are chosen as $b=a$ (accelerator and engine modes as depicted in Fig.~\ref{fig:1}(b)). In this setting, we can construct the Kraus decomposition of the resulting map from the quantum switch of channels $\mathcal{A}$ and $\mathcal{B}$ as described by the following operators
\begin{equation}
    K_{ij}:=M^b_{i}M^a_{j}\otimes\ket{0}_c\bra{0}+M^a_{j}M^b_{i}\otimes\ket{1}_c\bra{1},
\end{equation}
where $M^a_{j}$ and $M^b_{i}$ denote the Kraus operators acting on the working system for the channels $\mathcal{M}^a$ and $\mathcal{M}^b$, respectively, and the projectors $\ket{\ell}_c\bra{\ell}$ ($\ell=0,1$) lie in the quantum controller space. The resulting CPTP map for the quantum controlled channels has $16$ operators to be applied in the composite system-controller state %
\begin{equation}
   \rho^{\text{sw}}=\sum_{ij}K_{ij}(\rho^{(1)}\otimes\rho_{c})K_{ij}^{\dagger},
\end{equation}
where $\rho^{(1)}$ is the system initial Gibbs state and $\rho_{c}=\ket{c_{\theta}}\bra{c_{\theta}}$ is the controller initial state. 
We now have an interpolation controlled by  $0<\theta<\pi$ between the two measurement channels causal orders that are settled for $\theta=0$ (the natural order) and $\theta=\pi$ (the switched order), respectively. 

We note that, if the quantum controller is not observed, an incoherent mixture of the two causal orders of the measurement maps is obtained, 
\begin{equation} \label{Eqrhoswlocal}
   \rho^{\text{sw}}_{\text{inc}}= \text{tr}_{c}\left(\rho^{\text{sw}}\right)
   =\cos^2 \left(\frac{\theta}{2} \right) \rho^{(ab)}+\sin^2 \left(\frac{\theta}{2}\right)\rho^{(ba)},
\end{equation}
with $\rho^{(ab)}=\mathcal{M}^b\left(\mathcal{M}^a (\rho^{(1)})\right)$ (natural channels causal order) and $\rho^{(ba)}=\mathcal{M}^a\left(\mathcal{M}^b(\rho^{(1)})\right)$ (switched channels causal order). This scenario is depicted in Fig.~\ref{fig:3}(b). Moreover, by rewriting Eq.~(\ref{Eqrhoswlocal}) as $\rho^{\text{sw}}_{\text{inc}}=\frac{1}{2} {\mathbb{I}}+(a-1/2)\cos{(\theta)} \sigma_z$, it becomes clear that for the quantum controller angle $\theta=\pi/2$, we have a maximal mixture for any choice of the measurement parameter $a$, and the same happens when $a=1/2$ for any choice of the quantum controller angle $\theta$.

\begin{figure}
    \centering
    \includegraphics[width=1\columnwidth]{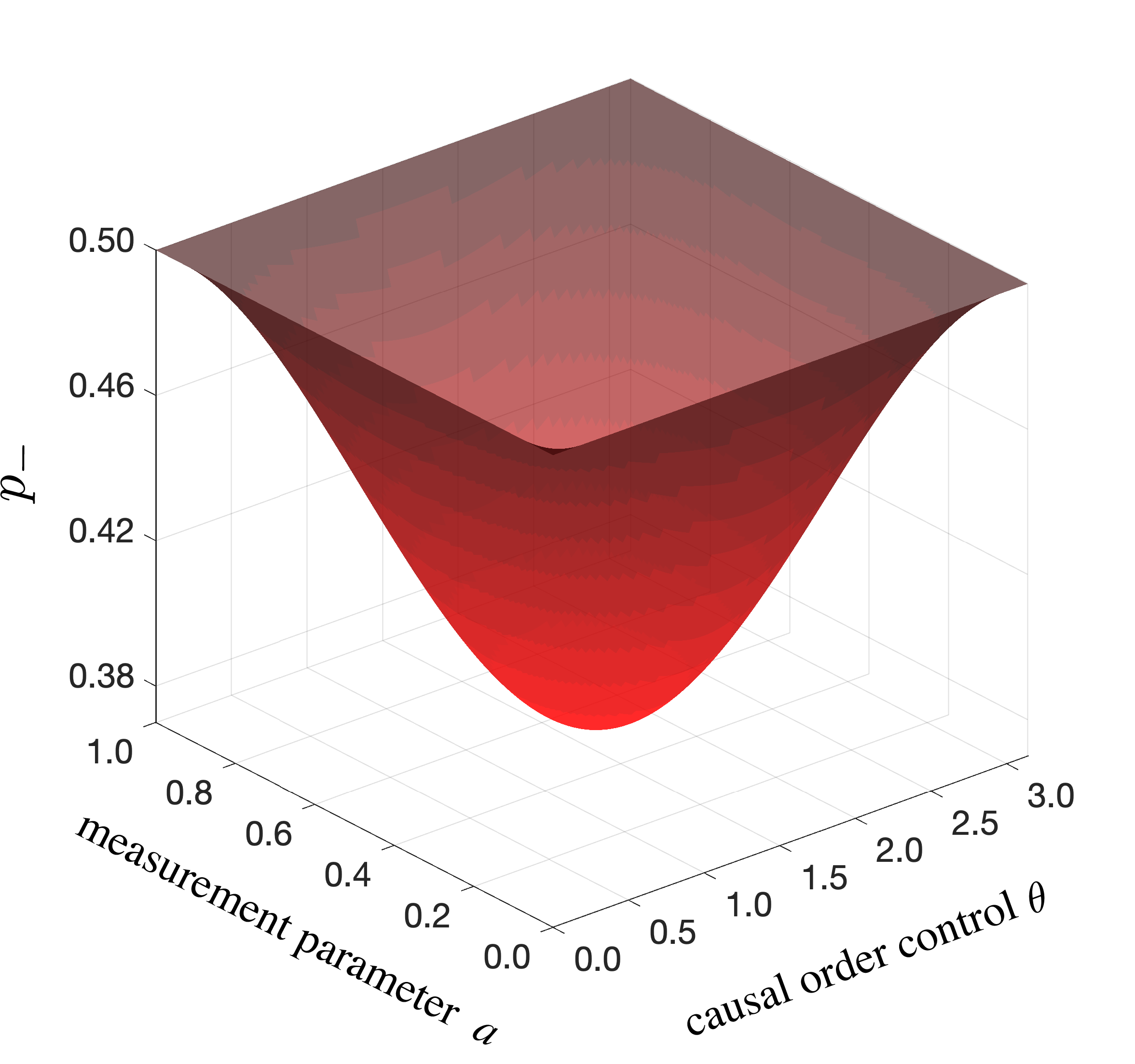}
    \caption{Probability for the projection of the causal order controller in the state $\ket{x_{-}}$. We note that, $p_{-}$ varies from $38\%$ to $50\%$.The probability for the projection of the causal order controller in the state $\ket{x_{+}}$ is complementary to the figure, since $p_{+}=1-p_{-}$.}
    \label{fig:4}
\end{figure}

Let us now, consider the indefinite causal order scenario by projecting the controller system in an orthogonal basis, $\ket{x_{\pm}}=(\ket{0}\pm\ket{1})/\sqrt{2}$. These two projective measurements on the causal order controller happen with probability $p_{{\pm}}=\left[1\pm a(1-a)\sin{\theta}\right]/2$ (displayed in Fig.~\ref{fig:4}). After the post-selection of the controller measurement, the working substance state turns out to be 
\begin{equation}
   \rho^{({\pm})}= \frac{1}{2 p_{\pm}} \rho^{\text{sw}}_{\text{inc}}+\left(1-\frac{1}{2 p_{\pm}}\right) \rho^{(1)}.
   \label{EqSW+-rho}
\end{equation}

The differences between the incoherent and coherent (with post-selection) control of causal orders are remarkable in the present context. First, when both measurement channels $\mathcal{M}^a$ and $\mathcal{M}^b$ are applied after the first thermalization stroke in each defined order (natural or switched), it completely erases the information of the initial equilibrium state, the states $\rho^{(ab)}$ and $\rho^{(ba)}$ (in Eq.~(\ref{Eqrhoswlocal})) do not carry any information about the cold environment temperature (although in the limits of the measurement parameter, these states depend on the initial thermal state). Second, when we project the controller on a state of an orthogonal basis ($\ket{x_\pm}$), it is clear from Eq.~(\ref{EqSW+-rho}), that we have an interpolation (as a function of the parameters $a$ and $\theta$) between the incoherent mixture of the two orders of the measurement maps (natural and switched) and the initial thermal state. Then, this partial erasing of the initial thermal state information can be interpreted as being the result of quantum interference effects between the two causal orders with a quantum control. In the following, we discuss how to use this quantum interference effect on the order controller to enhance the performance of a thermal device.

\section{Thermal devices powered by generalized measurements with indefinite causal order}

\subsection{Coherent controlled-device}
\begin{figure}
    \centering
    \includegraphics[width=1\columnwidth]{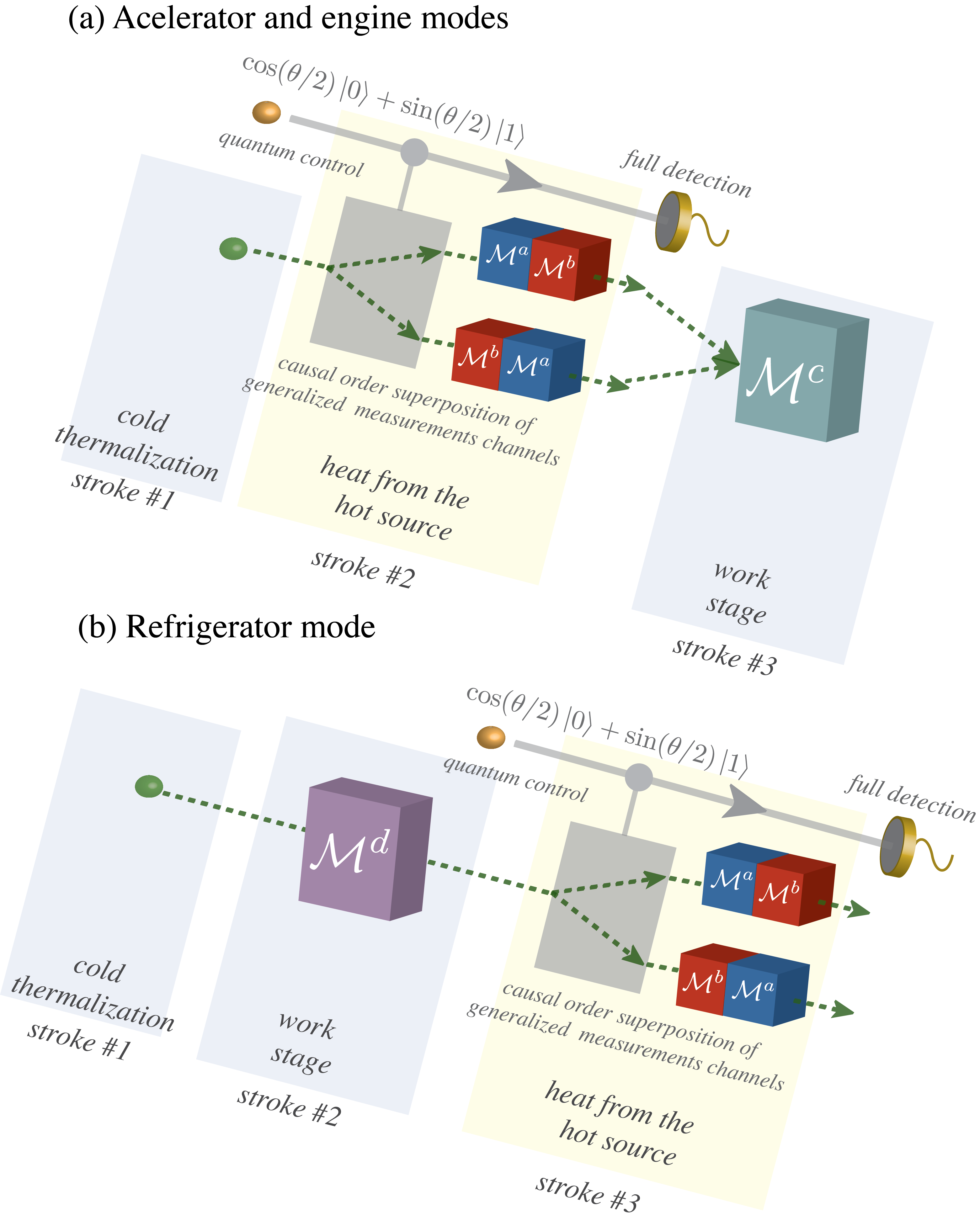}%
    \caption{ Quantum thermal device with indefinite causal order between measurement channels $\mathcal{M}^a$ and $\mathcal{M}^b$. (a) Accelerator and engine cycles. Starting from the cold thermal equilibrium state at inverse temperature $\beta$ (stroke 1), measurement channels ($\mathcal{M}^a$ and $\mathcal{M}^b$) with indefinite causal order are applied (stroke 2) transferring heat from the meters to the working substance. Next, the isentropic measurement channel $\mathcal{M}^c$ is applied to extract or invest work (stroke 3). (b) Refrigerator cycle. Starting from the cold thermal equilibrium state, the isentropic measurement channel $\mathcal{M}^d$ is applied to invest work (stroke 2). Next measurement channels ($\mathcal{M}^a$ and $\mathcal{M}^b$) with indefinite causal order are applied (stroke 3) to establish a heat flux from the cold environment to the meters.}
    \label{fig:5}
\end{figure}

\begin{figure*}[th]
    \centering
    \includegraphics[width=1\textwidth]{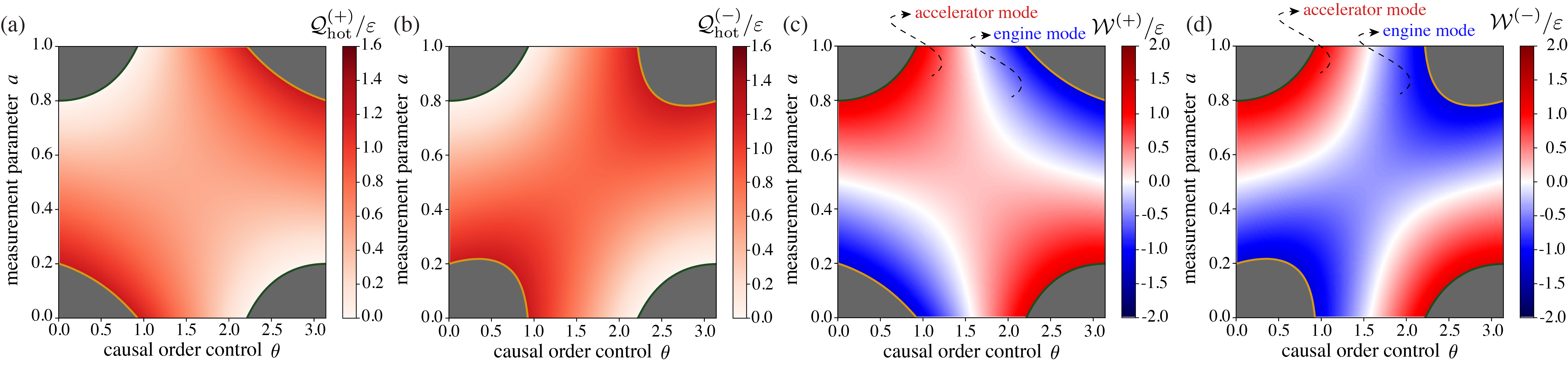}
    \caption{Heat and invested work for the quantum engine and accelerator with indefinite causal order of measurement channels. (a) Heat absorbed from the meters $\mathcal{A}$ and $\mathcal{B}$ as function of the measurement parameter $a$ ($b=a$) and the causal-order-control parameter $\theta$, considering the projection of the causal order controller in $\ket{x_{+}}$. (b) Absorbed heat considering the projection of the causal order controller in $\ket{x_{-}}$.
    (c) Work by the channel $\mathcal{M}^c$ considering the projection of the causal order controller in $\ket{x_{+}}$.  (d) Work considering the projection of the causal order controller in $\ket{x_{-}}$. The thermal device operates as an engine when $\mathcal{W}^{(\pm)} \leq0$ (blue color gradient) and it is a thermal accelerator when $\mathcal{W}^{(\pm)} > 0$ (red color gradient). The gray regions are excluded from the respective operation modes. All quantities were computed for $\beta \varepsilon = 1.39$.}
    \label{fig:6}
\end{figure*}

Here, we will explore the distinct operation modes of the measurement powered thermal device with indefinite causal order of the measurement channels $\mathcal{M}^a$ and $\mathcal{M}^b$, as illustrated in Fig.~\ref{fig:5}(a) and (b). Again in the first stroke, we consider a thermalization with the cold environment, initializing the system in the thermal equilibrium state $\rho^{(1)}$ at inverse temperature $\beta$. Now, the second stroke (for the accelerator and engine modes as depicted in Fig.~\ref{fig:5}(a)) is the quantum controlled switch of the measurement channels ($\mathcal{A}$ and $\mathcal{B}$) causal orders, as discussed in the previous section. Such a causal order will be controlled by the parameter $\theta$ and the controller state will be further post-selected, by a projective measurement, resulting in the state (\ref{EqSW+-rho}). 
In this case, the internal energy variation in the second stroke will be given by
\begin{equation}
  \expval{\Delta U^{(2),{(\pm)}}}=\frac{\varepsilon}{2p_{_{\pm}}}\left[ (1-2a)\cos{(\theta)}+\tanh(\beta\varepsilon)\right],
\end{equation}
where index $\pm$ refers to each possible outcome of the quantum controller selective measurement in the orthogonal basis, $\ket{x_\pm}$. The von Neumann entropy variation for the second stroke with indefinite causal order is $\Delta S^{(2),(\pm)} =S(\rho^{(\pm)})-S(\rho^{(1)})$ (where $\rho^{(\pm)}$ is defined in Eq.~(\ref{EqSW+-rho})). 
A not-null variation of the von Neumann entropy indicates heat flux to or from the meter, in this way the internal energy variation is associated to the heat transfer,  $\mathcal{Q}^{(\pm)}_{\text{hot}}=\expval{\Delta U^{(2),{(\pm)}}}$. As discussed before, an isentropic variation of energy can occur for the state in thermal equilibrium only when the following equality is satisfied $\expval{\Delta U^{(2),{(\pm)}}}=2\varepsilon\tanh(\beta\varepsilon)$. Let us introduce the following useful function,
\begin{equation}
   \Omega_\pm\equiv \frac{1}{4p_{\pm}}\left(1+\frac{1-2a}{\tanh(\beta \varepsilon)}\cos{(\theta)} \right),
\end{equation}
to evaluate the operation regimes of the quantum device. The condition, $\Omega_\pm = 1$, implies in an isentropic process ($\Delta S^{(2),(\pm)} =0$). At this point, it is noteworthy that for the choice of measurement parameters, $b=a$ (for the channels $\mathcal{A}$ and $\mathcal{B}$), a particular quantum controller of causal orders, $\theta=\pi/2$ (with maximum initial coherence), is not able to produce an isentropic variation of the internal energy in the second stroke with indefinite causal order, since there is no solution to reach the condition $\Omega_\pm = 1$. Another limit is to emphasize what happens when the switch operation does not change the initial thermal equilibrium state, which is associated to the condition $\Omega_\pm=0$. If the condition $0<\Omega_\pm<1$ is satisfied, the quantum thermal device with indefinite causal order can operate as an accelerator or an engine ($\mathcal{Q}^{(\pm)}_{\text{hot}}>0$ and $\Delta S^{(2),(\pm)}>0$). In Figs.~\ref{fig:6}(a) and \ref{fig:6}(b), we display the heat absorbed from the meters ($\mathcal{A}$ and $\mathcal{B}$) in the second stroke as a function of the measurement parameter $a$ (with $b=a$) and the causal-order-control parameter $\theta$, excluding the region out of the limit for $\Omega_\pm$ (out of the accelerator and engine operation mode).

Next, we consider an additional isentropic generalized measurement channel $\mathcal{M}^{(c)}$ to extract or invest work from or on the system in the stroke three, as illustrated in Fig.~\ref{fig:5}(a). The Kraus decomposition of the generalized measurement channel $\mathcal{M}^{(c)}$ will be the same of the channel $\mathcal{M}^{(b)}$ replacing the measurement parameter $b$ by $w_{\pm}$, explicitly given by $M^c_1=\sqrt{1-w_{\pm}}\ket{1}\bra{1}$, $M^c_2=\sqrt{1-w_{\pm}}\ket{1}\bra{0}$, $M^c_3=\sqrt{w_{\pm}}\ket{0}\bra{0}$ and $M^c_4=\sqrt{w_{\pm}}\ket{0}\bra{1}$. The isentropic non-selective measurement ($\Delta S^{(3),(\pm)}=0$) is performed adjusting the measurement parameter of the channel $\mathcal{C}$, accordingly with $p_{\pm}$ and $\theta$, as  
\begin{equation} \label{eq:parameterisenpm}
    \begin{aligned}
        w_\pm =  &\frac{1}{2p_{{\pm}}}\left[\frac{1}{2}- \left(a-\frac{1}{2} \right)\cos(\theta) \right]\\+ &\left( 1- \frac{1}{2 p_\pm}\right)\frac{1}{2}\left[1-\tanh(\beta\varepsilon)\right].
    \end{aligned}
\end{equation}
From this choice follows that the internal energy variation in this third stroke (performed by the measurement channel $\mathcal{M}^{c}$) will be given by
\begin{equation}\label{Workswitch}
\begin{aligned}
   \mathcal{W}^{(\pm)}&=\expval{ \Delta U^{(3),(\pm)} }
    = \text{tr}\left[H(\mathcal{M}^c(\rho^{{(\pm)}})-\rho^{{(\pm)}})\right] \\
    &=\frac{\varepsilon}{p_{{\pm}}}\left[ (2a-1)\cos(\theta)+(2p_{{\pm}}-1)\tanh(\beta \varepsilon)\right] ,
     \end{aligned}
\end{equation}
where $\mathcal{W}^{(\pm)}>0$ means work investment, or $\mathcal{W}^{(\pm)}<0$ is related to work extraction in the measurement channel $\mathcal{C}$. In Figs.~\ref{fig:6}(c) and \ref{fig:6}(d), we display the invested work to the meter $\mathcal{C}$ in the third stroke as a function of the measurement parameter $a$ (with $b=a$) and the causal-order-control parameter $\theta$, excluding the region out of the accelerator and engine operation mode. Negative values of $\mathcal{W}^{(\pm)}$ are associated with the engine operation mode and positive ones represent the thermal accelerator mode. In the interval $1/2<\Omega_\pm<1$ the device operates as an engine, whereas in the complementary interval $0<\Omega_\pm<1/2)$ the device is a thermal accelerator.

It is interesting to note that, for the measurement powered cycle with definite causal order presented in Sec.~II, no work can be invested or extracted with non-selective generalized measurements (in the third stroke) when we set $a=1/2$, which leads to $\rho^{(2)}=\mathcal{M}^a(\rho^{(1)})= {\mathbb{I}}/2$, a maximal mixture state.  When we consider an incoherent control of the causal order, the state just after the second stroke will be given by Eq.~(\ref{Eqrhoswlocal}) and if we set $a=1/2$ (and/or $\theta=\pi/2$) it turns out to be the maximal mixture state. So, in this case, no work can be invested or extracted with further non-selective generalized measurements (in the third stroke). Remarkably, for coherent control of the causal order (with a post-selection on an orthogonal basis), it is possible to invest or extract work (in the third stroke), when we set $a=1/2$ (or/and $\theta=\pi/2$), owing to quantum interference effects of the indefinite causal order of the measurement channels $\mathcal{A}$ and $\mathcal{B}$. This result can be observed in Figs.~\ref{fig:6}(c) and \ref{fig:6}(d).

\begin{figure*}[th]
    \centering
    \includegraphics[width=1\textwidth]{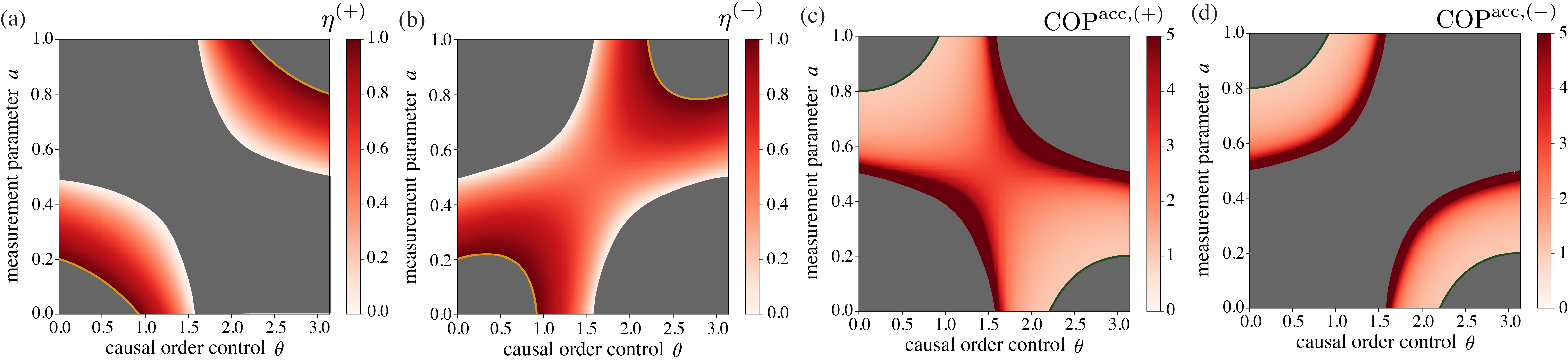}
    \caption{Engine efficiency and accelerator COP with indefinite causal order of measurement channels. (a) Efficiency in the engine operation mode, considering the projection of the causal order controller in $\ket{x_{+}}$. (b) Efficiency in the engine operation mode, considering the projection of the causal order controller in $\ket{x_{-}}$. (c) Coefficient of performance in the thermal accelerator mode, considering the projection of the causal order controller in $\ket{x_{+}}$. (d) Coefficient of performance in the thermal accelerator mode, considering the projection of the causal order controller in $\ket{x_{-}}$. The gray regions are excluded  from the respective operation modes. All quantities were computed for $\beta \varepsilon =1.39$.}
    \label{fig:7}
\end{figure*}

\begin{figure}
    \centering
    \includegraphics[width=1\columnwidth]{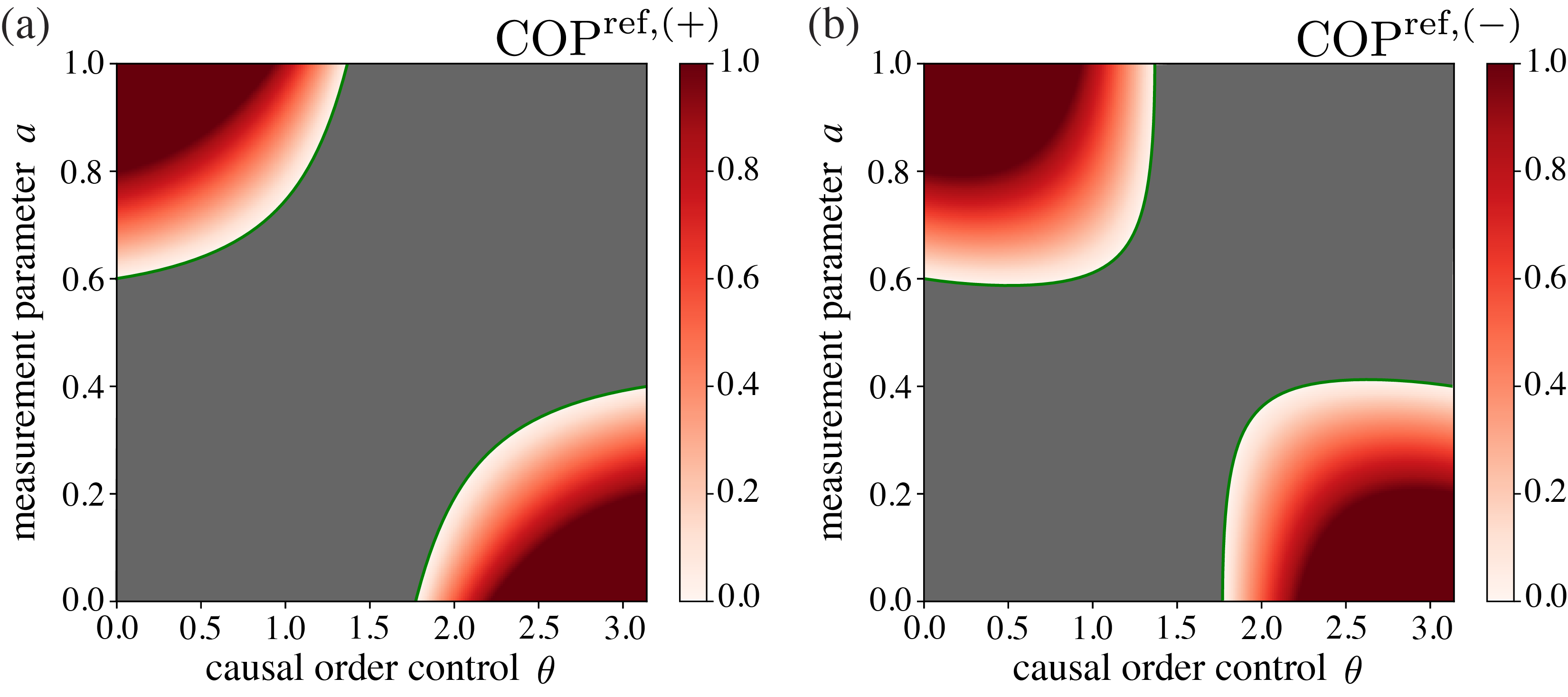}
    \caption{Refrigerator COP with indefinite causal order of measurement channels. (a) Coefficient of performance in the refrigerator mode, considering the projection of the causal order controller in $\ket{x_{+}}$. (b) Coefficient of performance in the refrigerator mode, considering the projection of the causal order controller in $\ket{x_{-}}$. The gray regions are excluded from the respective operation modes. Both quantities were computed for $\beta \varepsilon =0.45$.}
    \label{fig:8}
\end{figure} 

Finally, to close the cycle, the system interacts again with the cold environment and the internal energy variation is given by 
\begin{equation}
\begin{aligned}
\mathcal{Q}^{(\pm)}_{\text{cold}}
&=\expval{\Delta U^{(1),(\pm)}}=
\text{tr}\left[H(\rho^{(1)}-\mathcal{M}^c(\rho^{(\pm)})\right]\\
   &= -\frac{\varepsilon}{2p_{\pm}}\left[(2a-1)\cos(\theta)-(1-4p_{\pm})\tanh(\beta\varepsilon)\right].
    \end{aligned}
\end{equation}
In both the accelerator and engine operation modes $\mathcal{Q}^{(\pm)}_{\text{cold}}<0$.

The efficiency for the engine mode of the device with indefinite causal order can be written as
\begin{equation}
\label{eq:efc}
    \eta^{(\pm)}=2- \frac{1}{\Omega_\pm}.
\end{equation}
The coefficient of performance of the thermal accelerator mode turns out to be
\begin{equation}
    \text{COP}^{\text{acc},{(\pm)}}=1-\left(2-\frac{1}{\Omega_\pm}\right)^{-1}.
\end{equation}
The figures of merit for both operation modes thermal accelerator and engine are displayed in Figs.~\ref{fig:7}(a) to \ref{fig:7}(d), where we consider the projection of the causal order controller on the state $\ket{x_{\pm}}$. 

The operation of the quantum thermal device with indefinite causal order depends on the projective measurement of the order-controller qubit. In a cycle implementation, the generalized measurement channels parameters should be adjusted according to one of the possible outcomes, $\pm$, for the controller projection. In the case of an undesired measurement outcome, one should restart the aforementioned cycle before investing/extracting work, and repeat the switch operation until the desired projective result is obtained. In this sense, there is no need to include the probabilistic effect on the figures of merit for these both thermal modes (accelerator and engine). Of course, this probabilistic operation will affect the device power in contrast with the deterministic operation of the definite causal order device introduced in the Sec. II. The success probability, for the projection of the controller qubit in the state $\ket{x_{-}}$, in our example, varies from about  $38\%$ to  $50\%$ as displayed in Fig.~\ref{fig:4}.

It is also possible to set the generalized measurement powered thermal device with indefinite causal order to operates in the refrigerator mode. To this end, we will consider in the second stroke the isentropic generalized measurement channel $\mathcal{M}^d$ and the indefinite causal order of channels $\mathcal{M}^a$ and $\mathcal{M}^b$ in the third stroke, as illustrated in Fig.~\ref{fig:5}(b). The Kraus decomposition of the generalized measurement channel $\mathcal{M}^{(d)}$ will be the same of the channel $\mathcal{M}^{(a)}$ replacing the measurement parameter $a$ by $d$, explicitly given by $M^d_1=\sqrt{1-d}\ket{0}\bra{0}$, $M^d_2=\sqrt{1-d}\ket{0}\bra{1}$, $M^d_3=\sqrt{d}\ket{1}\bra{1}$, and $M^d_4=\sqrt{d}\ket{1}\bra{0}$. Setting $d=\left(1+\tanh(\beta \varepsilon)\right)/2$, we ensure that $\Delta S^{(2)}=0$ and the invested work in this stroke will be given by 
\begin{equation}
\mathcal{W}_{\text{inv}}=\expval{\Delta U^{(2)}}=2\varepsilon\tanh(\beta \varepsilon)>0.
\end{equation}

In next stages a heat flow from the cold environment to the meters $\mathcal{A}$ and $\mathcal{B}$ will be stabilised with the indefinite causal order of later measurement channels. In this case the internal energy variation in the third stroke will be written as
\begin{equation}
\begin{aligned}
    \mathcal{Q}^{(\pm)}_{\text{hot}}&=\expval{\Delta U^{(3),{(\pm)}}} \\ &=-\frac{\varepsilon}{2p_{\pm}}\left[(2a-1)\cos(\theta)+\tanh(\beta\varepsilon))\right],
    \end{aligned}
\end{equation}
where we have considered the following $\Omega_\pm<0$, which leads to $\mathcal{Q}^{(\pm)}_{\text{hot}}<0$ heat delivered to the meter.

\begin{figure*}[th]
    \centering
    \includegraphics[width=1\textwidth]{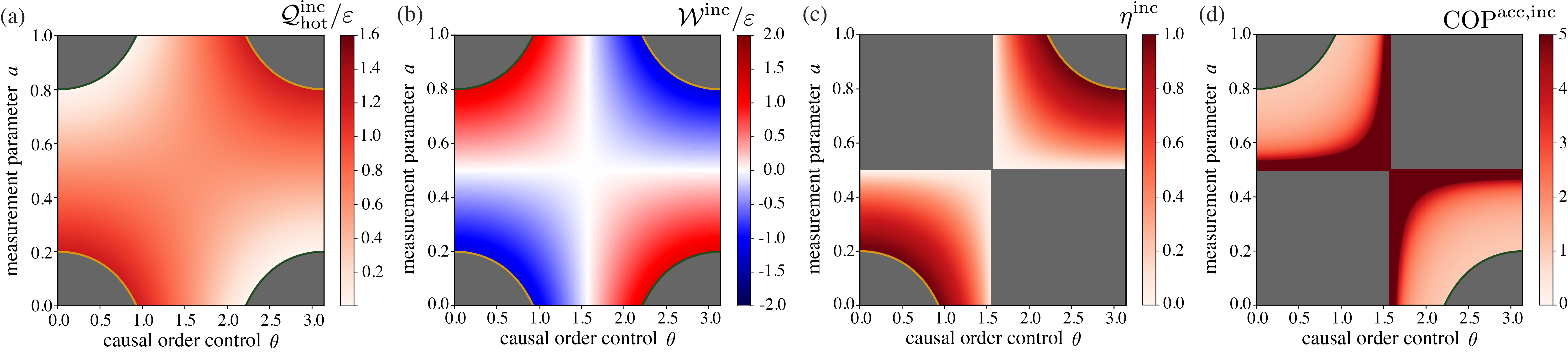}
    \caption{Thermal device with incoherent causal order control. (a) Heat absorbed from the meters $\mathcal{A}$ and $\mathcal{B}$ as a function of parameters $a$ and $\theta$.
    (b) Work invested by the channel $\mathcal{M}^c$. (c) Efficiency in the engine mode. (d) Coefficient of performance in the accelerator mode. The gray regions are excluded from the respective operation modes. All quantities were computed considering the scenario where the causal order controller is not observed (incoherent control) and for $\beta \varepsilon = 1.39$.}
    \label{fig:9}
\end{figure*} 

Finally, the thermalization with cold environment closes the cycle, resulting in the heat exchange with the cold reservoir as 
\begin{equation}
\begin{aligned}
    \mathcal{Q^{(\pm)}_{\text{cold}}}&=\expval{\Delta U^{(1),(\pm)}} \\
    &= \frac{\varepsilon}{2p_{{\pm}}}\left[ (2a-1) \cos(\theta) +(1-4p_{{\pm}})\tanh(\beta \varepsilon) \right].
    \end{aligned}
\end{equation}
The device will absorb heat from the cold environment ($\mathcal{Q^{\pm}_{\text{cold}}})>0$. 

The coefficient of the performance of the refrigerator setup with indefinite causal order is 
\begin{equation} 
        \text{COP}^{\text{ref},(\pm)}=\frac{\mathcal{Q}^{(\pm)}_{\text{cold}}}{\mathcal{W}_{\text{inv}}}=\frac{1}{2p_{{\pm}}}-(\Omega_\pm+1).
\end{equation}
In Fig.~\ref{fig:8}, we display the $\text{COP}^{\text{ref},(\pm)}$ considering the projection of the causal order controller in state $\ket{x_{\pm}}$.

The refrigerator powered by a generalized measurement with indefinite causal order also depends on the projective measurement on the order-controller qubit. The probability of success, in this case, is also displayed in Fig.~\ref{fig:4}. In the case of an undesired measurement outcome, the work invested in the second stroke can be completely recovered by a subsequent isentropic measurement channel (work channel) and the protocol should be repeated to get the desired projection on the order-controller qubit. Of course, the probabilistic nature of the refrigerator protocol with indefinite causal order will affect the cooling rate.

\subsection{Incoherent controlled-device}

To highlight the differences between applying a coherent or incoherent quantum control to perform the order switch of the measurement channels (i.e., the differences between having quantum interference effects or just a stochastic mixture of the orders), we now analyze the incoherent mixture of the two causal orders of the measurement maps ($\mathcal{A}$ and $\mathcal{B}$) and compare it with the engine or accelerator modes developed with coherent control. In this case, the quantum controller is not observed as depicted in Fig.~\ref{fig:3}(b). The incoherent switch of the causal order will be employed in the second stroke, leading the working substance to the state described in Eq.~(\ref{Eqrhoswlocal}). Now, we introduce the following useful function, to analyze the regimes in which the thermal device with an incoherent mixture of orders operates,
\begin{equation}
   \Omega_{\text{inc}}\equiv \frac{1}{2}\left(1+\frac{1-2a}{\tanh(\beta \varepsilon)}\cos{(\theta)} \right).
\end{equation}

Let us evaluate the scenario, where the incoherent switch of casual orders of the measurement maps $\mathcal{A}$ and $\mathcal{B}$ occurs in the second stroke. Considering the interval, $0<\Omega_{\text{inc}}<1$, the heat absorbed from the meters ($\mathcal{Q}^{\text{inc}}_{\text{hot}}>0$) can be written as
\begin{equation}
  \mathcal{Q}^{\text{inc}}_{\text{hot}}= \varepsilon \left[ (1-2a)\cos (\theta) +\tanh (\beta \varepsilon) \right].
\end{equation}
In this respect, the von Neumann entropy variation is positive, $\Delta S^{(2), \text{inc}}>0$.
In Fig.~\ref{fig:9}(a), we display the absorbed heat in the second stroke with incoherent causal order control.

Once again, we consider an additional isentropic generalized measurement channel $\mathcal{M}^{(c)}$ to extract or invest work from or on the system in the stroke three. In the scenario with incoherent causal order control, the measurement parameter for the channel $\mathcal{M}^{(c)}$ is
\begin{equation}
    w_{\text{inc}} = \left[\frac{1}{2}-\left(a-\frac{1}{2}\right)\cos(\theta)\right],
\end{equation} 
while the Kraus decomposition has the form introduced in the previous section replacing $w_{\pm}$ by $w_{\text{inc}}$.
This choice turns the measurement channel $\mathcal{C}$ into an isentropic non-selective measurement for the incoherent controlled-device, so the the work (in the third stroke) may be written as 
\begin{equation}\label{Workinc}
   \mathcal{W}^{\text{inc}} = 2 \varepsilon \left[ (2a-1)\cos (\theta) \right].
\end{equation}
For the negative values of work (blue region in Fig.~\ref{fig:9}(b)), which happens for $\frac{1}{2}<\Omega_{\text{inc}}<1$, the device operates as an engine with efficiency
\begin{equation}
\label{eq:efinc}
    \begin{aligned}
           \eta^{\text{inc}}&={2 \left[ 1+\frac{\tanh (\beta \varepsilon)}{(1-2a)\cos (\theta)}\right]}^{-1}\\
           &=2-\frac{1}{\Omega_{\text{inc}}}.
    \end{aligned}
\end{equation}
While for positive values of work (red region in Fig.~\ref{fig:9}(b)), which happens for $0<\Omega_{\text{inc}}<\frac{1}{2}$, the device operates as an thermal accelerator with coefficient of performance
\begin{equation}
   \begin{aligned}
          \text{COP}^{\text{acc,inc}}&=\frac{1}{2} \left[ 1-\frac{\tanh (\beta \varepsilon)}{(1-2a)\cos (\theta)}\right]\\
          &=1-\left(2-\frac{1}{\Omega_{\text{inc}}}\right)^{-1}.
   \end{aligned}
\end{equation}

Figures \ref{fig:9}(c) and \ref{fig:9}(d) display the engine efficiency and the accelerator COP for the thermal device with incoherent causal order control. In this scenario, the thermal device operates in a deterministic way, since the causal-order-control qubit is not observed. The differences between coherent and incoherent causal order control are prominent and can be straightforwardly observed by comparing Figs.~\ref{fig:6}(c) and \ref{fig:6}(d) with Fig.~\ref{fig:9}(b). In particular, we notice that when we set the measurement parameter as $a = 1/2$ or the causal order parameter as
$\theta  = \pi/2$, it is not possible to invest or extract
work (in the third stroke) in the incoherent case (Fig.~\ref{fig:9}(b)). On the order hand, in the coherent case it turns out to be a possible work extraction or investment due to quantum interference effects of the indefinite causal order for the measurement channels (Figs.~\ref{fig:6}(c) and \ref{fig:6}(d)).

Finally, employing the auxiliary functions $\Omega_{\pm}$ and $\Omega_{\text{inc}}$, one can easily demonstrate for a fixed parameters pair $(a,\theta)$, where it is possible to obtain an advantage by employing coherent control over its incoherent version in the thermal device. For instance, consider the engine efficiencies for the two versions of the order controller given by Eqs.~(\ref{eq:efc}) and (\ref{eq:efinc}). For a fixed pair of parameters, it is straightforward to verify that the advantage for work extraction occurs whenever the condition $\Omega_{\pm}>\Omega_{\text{inc}}$ is satisfied, which is associated with $p_{\pm}<\frac{1}{2}$. Therefore, as can be observed in Fig.~\ref{fig:4}, the probability for the projection on $\ket{x_{-}}$ of the quantum controlled-device, enhances the heat engine performance for all the points in operation mode except where $p_{-}=\frac{1}{2}$. Moreover, the maximum advantage (with coherent control) occurs for the two values in which the incoherent version has always a null efficiency, i.e, for $a=\frac{1}{2}$ and $\theta=\frac{\pi}{2}$ (as observed comparing Figs.~\ref{fig:7}(b) and \ref{fig:9}(c)). The same argument can be applied for the projection in $\ket{x_{+}}$, which can fuel the performance of the accelerator mode by employing measurements with indefinite causal order to enhance the natural heat flux (from the meters to the cold environment) with lesser invested work in relation to the incoherent mixture of the measurement orders.

%
%

%

%

\section{Conclusion}
\label{dis}

We introduced a quantum thermal device model powered by non-selective generalized measurements. Setting the measurement parameter, the device can operate as a thermal accelerator (transferring heat from the meter to a cold environment), an engine (extracting work to the meter), or a refrigerator (transferring heat from the cold environment to the meter). In this context, we discussed the impossibility of extracting work using a non-selective generalized measurement directly from an equilibrium thermal state and highlighted the need for a set of non-selective measurement channels (that lead to a non-equilibrium state in the intermediate stages) for such a task. We also investigated the implication of an indefinite causal order in the measurement channels, controlled by an ancillary quantum system. We explored the notion of a quantum switch of the measurement channels to extend the operation regimes of the device. As can be observed in the figures of merit for the devices with indefinite causal order of the measurement channels, by comparing the order control parameter, set as $\theta=0$ (without order switch) and $0<\theta<\pi$ (with order switch). Such performance behavior is associated with interference effects given by the indefinite causal orders which can not be obtained for the specific values of measurement parameters and the initial coherence of the quantum controlled device in both definite orders of measurement application and also on the incoherent mixture of them. Moreover, we also demonstrated that the coherent control over the order switcher enhances the device performance for some  parameters settings over the incoherent case.

The superposition principle of quantum mechanics offers the possibility of controlling operations beyond classical capabilities. In particular, the quantum switch of operations with coherent control has been claimed to
present a genuine superposition of causal orders.  In particular, the interpretation of such proposals as realizations of a process with indefinite causal structure has recently been questioned~\cite{paunkovic2020causal,ormrod2022causal,vilasini2022embedding}. In fact, there is an interesting and ongoing debate concerning causal structures, with counterpoints to some objections~\cite{hamette2022quantum}.
In Ref.~\cite{paunkovic2020causal}, the authors argued that in contrast to the gravitational switch, the current experimental implementations do not feature superposition of spacetime causal orders and that they are variants of the time double slit experiment~\cite{paunkovic2020causal}. Another recent investigation can be found in Ref.~\cite{vilasini2022embedding}, where it was argued that any physical implementation of an indefinite order process can ultimately be fine-grained to one that admits a fixed acyclic information-theoretic causal
order, that is compatible with the spacetime causal order. On the other hand, a counterpoint to the discussion is presented in Ref.~\cite{hamette2022quantum}, the authors introduce an unambiguous definition of causal order between two events in terms of the proper time of a third particle. From such a definition different notions of indefinite causal order (information-theoretic and gravitational) were connected. A superposition of causal order that cannot be rendered definite was also explored in Ref.~\cite{hamette2022quantum}.

Regardless of the discussion associated with the causal order interpretation, the quantum control of operations in thermodynamic cycles leads to new effects that can not be reproduced with incoherent control of the same operations~\cite{felce2020quantum,cao2022quantum,nie2022experimental,simonov2022work}. Indeed, our results corroborate that there are effects for work extraction in the measurement-powered quantum thermal devices that are observed with indefinite causal order and can not be observed using the same operations with a definite order.  A similar conclusion can be seen regarding activation of work extraction when quantum maps are applied with indefinite causal order using finite-time regime protocols~\cite{simonov2022work}. By employing the notion of ergotropy, which quantifies the maximal work that can be extracted from a quantum system through a cyclic unitary transformation of the reference Hamiltonians parameters, gain was observed in ergotropy with the quantum switch of the maps compared to their consecutive application in Ref.~\cite{simonov2022work}. Further developments in this context, would be associated with exploring non-symmetrical measurement parameters on the quantum switch operation and the role of an initial coherence in the system due to a not complete thermalization (with a cold environment) in a finite-time regime.

\section*{Acknowledgments}
The authors acknowledge Ismael Paiva for his helpful suggestions. The authors acknowledge financial support from the Federal University of ABC, CAPES, CNPq, FAPESP, and the National Institute for Science and Technology of Quantum Information (CNPq, INCT-IQ 465469/2014-0). P.R.D. acknowledges support from the Foundation for Polish Science (IRAP project, ICTQT, contract no. MAB/2018/5, co-financed by EU within Smart Growth Operational Programme) and CAPES Research Grant No. 88887.354951/2019-00. R.M.S. also acknowledges Ministry of Science and Technology (China), trough the High-End Foreign Expert Program (grant n.~G2021016021L).


\bibliography{citations}

\end{document}